\documentclass{iau} 
\usepackage{graphicx}

\title[IAU-S322~~Galactic Center compared with extragalactic nuclei] 
{The Galactic Center compared with nuclei of nearby galaxies}

\author[F. Combes]   
{Francoise Combes$^1$}

\affiliation{$^1$Observatoire de Paris, LERMA, College de France, CNRS, PSL, Sorbonne Univ.
UPMC, F-75014, Paris, France \\ email: {\tt francoise.combes@obspm.fr}} 

\pubyear{2016}
\volume{322}  
\jname{The Multi-Messenger Astrophysics of the Galactic Centre}
\editors{Steve Longmore, Geoff Bicknell \& Roland Crocker, eds.}
\begin{document}

\maketitle

\begin{abstract}
Understanding our Galactic Center is easier with insights
from nearby galactic nuclei. Both the star formation activity
in nuclear gas disks, driven by bars and nuclear bars,
and the fueling of low-luminosity AGN, followed by
feedback of jets, driving molecular outflows, were certainly present
in our Galactic Center, which now appears to be quenched.
Comparisons and diagnostics are reviewed, in particular of $m=2$ and
$m=1$ modes, lopsidedness, different disk orientations, and fossil
evidences of activity and feedback.
\keywords{Galaxy: center --  Galaxy: kinematics and dynamics  --  Galaxy: nucleus  
-- galaxies: active -- galaxies: jets  }
\end{abstract}

\firstsection 
\section{Introduction}

It is now well established that our Galactic center is very quiet, both
on the point of view of star formation, and of nuclear activity.  The
luminosity of Sagittarius A* is only 10$^{-9}$ times the Eddington luminosity
(or 300 L$_\odot$) (e.g. Genzel et al 2010), and the star formation rate is
10 times lower than expected from the high molecular
surface density in the CMZ (Central Molecular Zone), e.g. Longmore et al. (2013). 
The reason for this low activity, and low efficiency of star formation might
be transient, since there is evidence of recent past activity (e.g. 
Ponti et al. 2010; Carretti et al. 2013). Our Galactic center appears to have been 
quenched by some feedback effect, which has raised the turbulence of the gas to a 
high level. The velocity dispersion, the radiation field and magnetic field are all
very high in this environment (Ferri\`ere et al 2007), and these might be clues for
the recent quenching.

The Galaxy is a common barred spiral, with strong non-axisymmetries, both
of $m=2$ and $m=1$ types, and these are powerful engines to fuel gas to the nuclear 
regions. The gas is present, and certainly other activity episodes are
expected in the future.

\section{Bars in galaxies to fuel AGN and star formation activities}

\begin{figure}[t]
\begin{center}
 \includegraphics[width=10cm]{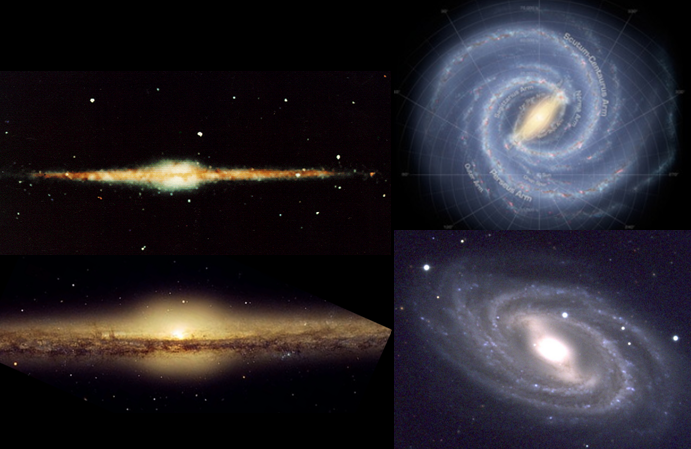} 
 \caption{Comparison between the Milky Way and nearby galaxies with similar morphology and dynamics:
on the left panel, the COBE satellite near-infrared image is compared to the edge-on spiral
NGC 4565, showing a boxy-peanut bulge; on the right panel, the artist view of the face-on
Milky Way is compared to the barred spiral NGC 3992.}   
   \label{fig1}
\end{center}
\end{figure}

 As shown in Figure \ref{fig1}, our Galaxy has a box/peanut shape bulge,
which is attributed to the vertical resonance of the bar (e.g. Combes \&
Sanders 1981, Ness et al. 2012), and a strong bar inside a spiral structure,
which is evolved dynamically, so that its pattern speed has relaxed to a low
value, allowing two Lindblad resonances (e.g. Binney et al 1991, 
Rodriguez-Fernandez \& Combes 2008). It is also possible that the weakening 
of the bar, due to the perpendicular orbits inside the two ILRs, and
the bulge thickening towards the center, has or will lead
to the decoupling of a secondary bar, a faster embedded bar (e.g. Friedli
\& Martinet 1993, Alard 2001). This nuclear bar would exist at $\sim$ 100pc scale.

These dynamical features help to understand the gas flows in the Galaxy, and the 
possible formation of gaseous rings, like the CMZ at radii $\sim$150-200pc. 
It can be shown that the bar exerts torques on the gaseous disk, and the sign of the torques
change at each resonance. The torques are negative inside corotation, and can drive the gas towards
the inner Lindblad resonance (ILR), where it accumulates in a ring. For usual rotation curves,
where the rate of orbit precession inside the ILR ($\Omega-\kappa/2$) increases with radius, 
spiral arms lead inside the ILR, and the torques are negative, so that the gas is
stalled. It has to await viscous torques to infall. When the central region is under the 
gravitational influence of the central black hole, the precession rate decreases with radius,
and the spirals are trailing, which reverses the sign of the torques, and the gas is driven in.
This is a case recently observed with ALMA in NGC 1566 (Combes et al 2014).

 Torques have been computed in more than 20 nearby galaxies showing low-luminosity
nuclear activity (Seyfert, Liners), with gaseous maps at 
high interferometric resolution, and forces computed with red images at HST resolution
(e.g. Garcia-Burillo \& Combes 2012, NUGA project). Suprisingly, only one third
of galaxies show gas accretion at $\sim$ 100pc scale, due to a nuclear bar, or no ILR
in the center. In two thirds of the sample, the gas is stalled in a nuclear ring, or
driven outwards by the gravity torques. Given that all these galaxies should have
accreted gas at some step of their bar cycle, this result could be interpreted
as a typical galaxy like the Milky Way experiencung gas accretion at that scale only one 
third of its time.

\section{Off-centering, lopsidedness and the example of M31}

Closer to the center, and under the influence of the black hole (BH), orbits become
quasi-keplerian. With some self-gravity, this can trigger special modes
of densiy waves, with $m=1$ symmetry, provoking an off-centering of the central mass.
This mode allows the inner disk to lose angular momentum, and the gas to
fall onto the central BH (cf Reichard et al. 2009).
This $m=1$ mode appears clearly in the center of our neighbor M31, which shows a double nucleus:
two stellar components P1 and P2, P1 being the brightest. 
In fact, the P1-P2 ensemble is part of the same disk, which is lopsided and off-centered
through the $m=1$ mode (Bacon et al 2001). The BH mass is 30 times higher than in our
Galaxy, and the zone of influence wider. The nuclear stellar disk has a typical size of 
10pc in diameter. The eccentric disk model was shown to reproduce the observations,
both by a mass-less experiment (Peiris \& Tremaine 2003), and a self-consistent
N-body model (Bacon et al 2001). The $m=1$ pattern speed is 3km/s/pc, and the life-time of the 
wave is at least 3000 rotations. This configuration was shown to be able to drive the
angular momentum away (Saha \& Jog 2014).

\begin{figure}[t]
\begin{center}
 \includegraphics[width=13cm]{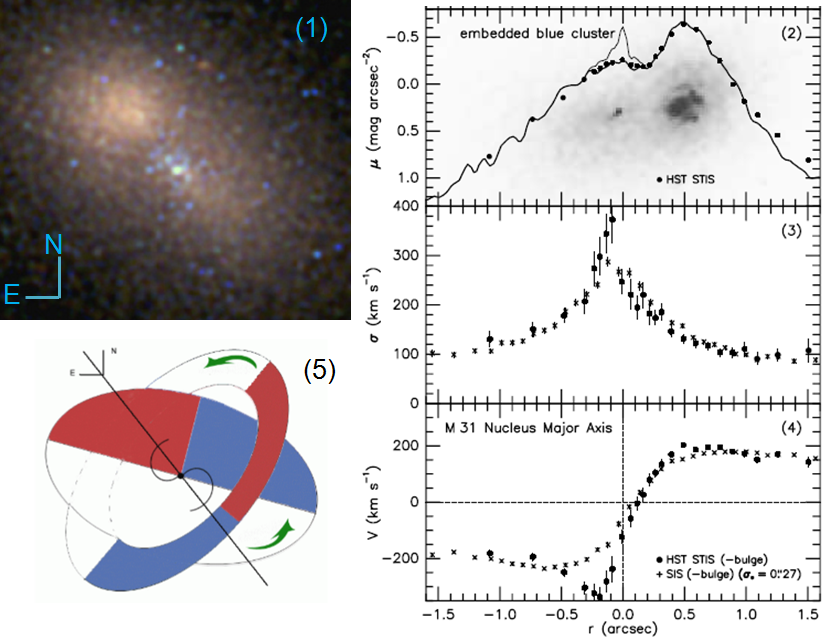}
 \caption{The nucleus of the most nearby spiral galaxy M31 (Andromeda) show features very similar
to the Galactic center: (1) the nuclear stellar structure can be decomposed in three components: 
P1 and P2 with yellow/red color (P1 being the brightest at NE), and P3 closed to the very center, i.e.
the supermassive black-hole, with blue color (from Lauer et al. 2012); 
(2) the radial profile of stellar surface density in red
light (HST, thick line), and the peak of P3 visible only in ultraviolet (light line); (3) and (4) 
show the velocity dispersion and the radial velocity profiles, from Bender et al. (2005).
The last panel (5) is a schematic interpretation of he molecular gas distribution, in different
disks (Melchior \& Combes 2011).}
   \label{fig2}
\end{center}
\end{figure}

More recently, it was shown in the ultraviolet that a third stellar component, P3,
dominates close to the BH. Contrary to P1-P2 which is an old-stellar disk, P2 is a blue 
young population, corresponding to a 200 Myr old starburst. Figure \ref{fig2}
shows the light distribution along the major axis, together with the velocity
dispersion, indicating that P3 corresponds to the BH position, and the asymmetric
velocity profile. P3 is a small stellar cluster, of typical size 0.4pc, with a different
orientation (inclination, position angle) from the P1-P2 disk. Like in our
Milky Way, showing at least two nuclear stellar disks with two different orientations.
 As far as the molecular gas is concerned, two different disks are also observed
(Melchior \& Combes 2011).

The existence of a 10Myr young stellar disk (of size 0.4pc) 
 in our Milky Way has raised the paradox of youth
(e.g. Genzel et al 2010). To form stars, a gas cloud is required to be dense enough
to resist the strong tidal force, which leads to 6 10$^{10}$ cm$^{-3}$ at a radius
of 0.1pc, but the gas observed is far from this density. This paradox has led to several
hypotheses, like inwards migration of the star cluster after formation far from the center,
or stars rejuvenated by collisions. But the most likely is the formation in a dense 
disk, in situ.

For M31, the same paradox is raised. How could the 200Myr young P3 cluster have formed?
The stellar cluster has a surprisingly high compactness (radius 1pc), and cannot come
only from BH-stripped giants. It must be possible to form stars in the strong
tidal field of the 10$^8$ M$_\odot$ BH.  The
migration scenario is even less likely than in the GC, since the mass
of the BH is much higher.

A scenario has been proposed by Chang et al. (2007). Some gas is released by 
stellar mass loss from the P1+P2 disk. The latter is experiencing
an $m=1$ mode, with a pattern speed $\sim$ 3-10 km/s/pc. This
fixes the size of P3. Outside a certain radius, the gas clouds participating in the $m=1$
wave are located on crossing orbits, which creates 
dissipation and infall, until a radius of $\sim$ 1pc. The radius of P3 is thus the
last non-crossing orbit. The gas mass available for P3 is about 10$^5$ M$_\odot$,
compatible with the mass loss from a disk P1+P2 of 10$^7$ M$_\odot$.
The stellar mass loss rate of 10$^{-4}$ M$_\odot$/yr fills P3 in about 500Myr. This
scenario should produce repeated star bursts, at that frequency.

\begin{figure}[t]
\begin{center}
 \includegraphics[width=13cm]{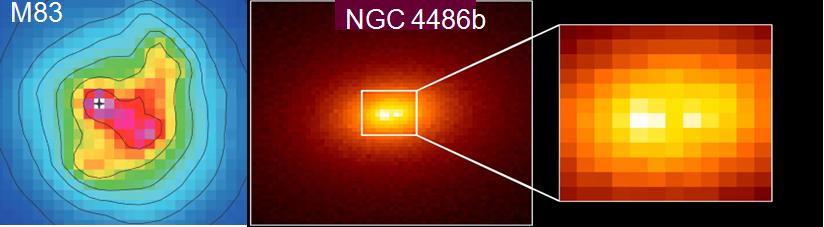}
 \caption{ {\bf Left:}  The nucleus of Messier 83, seen in the near-infrared
at 3.6 $\mu$m, showing a pronounced lopsidedness (Knapen et al. 2010).
{\bf Right:} HST image of NGC 4486b, showing a double nucleus similar to M31 (Lauer et al. 1996).}
   \label{fig3}
\end{center}
\end{figure}

The $m=1$ wave and off-centring exists also in our Galaxy. The well-known asymmetry
of the molecular parallelogram in the l-v diagram means that three quarters of the gas mass is at positive
longitude, and 1/4th at negative longitude.
The phenomenon of a lopsided disk is also frequently observed in external galaxies,
when there is sufficient spatial resolution: for instance M83 and NGC 4486b (see Figure \ref{fig3},
Thatte et al, 2000; Knapen et al 2010, Lauer et al 1996) or also VCC128, showing a double
nucleus (Debattista et al 2006). Some double nuclei, like in NGC 4654, are made of two distinct nuclear 
star clusters (Georgiev \& B\"oker 2014).

\section{Disks with different orientations}

\begin{figure}[t]
\begin{center}
 \includegraphics[width=13cm]{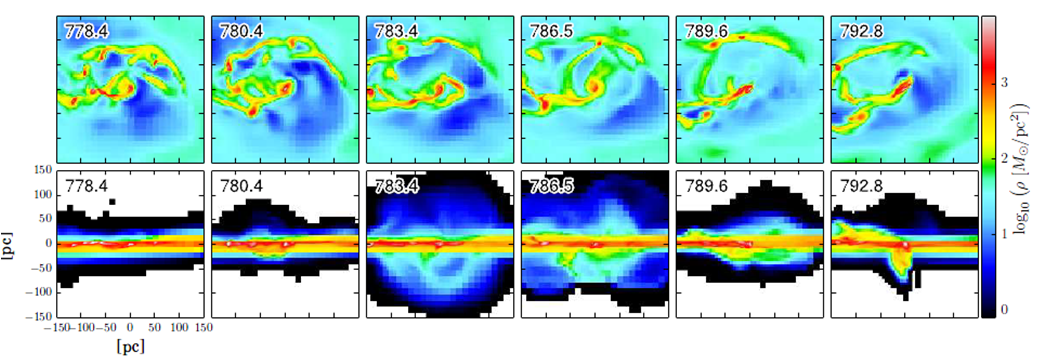} 
 \caption{N-body and hydrodynamical simulations of a Milky-Way like galaxy, developing
a bar, and showing gas inflow towards the center. When the gas has sufficiently accumulated,
the consequent starburst and feedback associated have ejected gas above the plane, which
falls back in a fountain. It settles in a polar plane, which may explain the various
orientation of gas and stellar disks in galaxy nuclei (from Renaud et al. 2015 and Emsellem
et al. 2015). The top row shows 6 snapshots face-on, and the bottom
row edge-on, from 778 Myr to 792 Myr.}
   \label{fig4}
\end{center}
\end{figure}

In M31 as well as in our Galaxy, nuclear disks have a different orientation than
the main disk, and also there can exist several different nuclear disks. How can 
they form?  During a high-resolution simulation of a Milky-Way-like galaxy,
such a phenomenon has been observed (see Figure \ref{fig4}). The bar gravity torques
progressively drive the gas inwards. After some time, the gas accumulation in the center
triggers a mini-starburst, and the associated supernovae feedback ejects some gas 
perpendicular to the plane. The latter can fall down in random directions, with different
orientations. In the simulation, the gas settled in a polar ring, the only orientation
stable with respect to differential precession. Certainly many other orientation
can occur in the real world.

\begin{figure}[t]
\begin{center}
 \includegraphics[width=12cm]{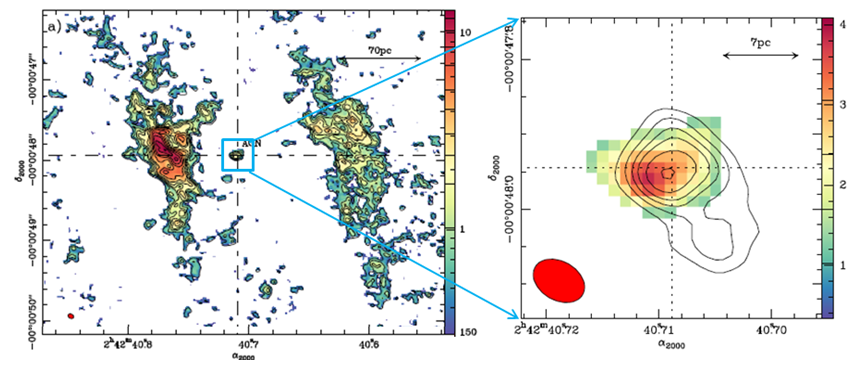} 
 \caption{ {\bf Left:} ALMA CO(6-5) map of the nucleus of NGC 1068, showing the circum-nuclear disk (CND)
of $\sim$300pc scale, and the off-centered molecular torus around the AGN. 
{\bf Right:} Zoomed view of the torus, with the CO(6-5) emission in colors, and the dust emission 
in contours. From Garcia-Burillo et al (2016).}
   \label{fig5}
\end{center}
\end{figure}

The non-alignment of nuclear disks with the host disk is frequent, as
shown by radio jets which are not perpendicular to their main disks
(e.g. Schmitt \& Kinney 2002, Jog \& Combes 2009). In NGC 4258, the maser disk of 0.2pc in size, 
is misaligned by 119$^\circ$  from its galaxy disk, and the radio jet grazes the plane.
This is also the case in NGC 1068 (Garcia-Burillo et al. 2014).
 The spatial resolution brought by the ALMA interferometer allows us to distinguish
molecular tori in nearby galaxies. In NGC 1068, the CO(3-2) torus appears
more inclined than the water maser disk, and subject to the Papaloizou-Pringle instability
(see Figure \ref{fig5}).

\section{Intermittent feedback and quenching}

 The fueling of gas towards the inner regions has been studied
through zooming re-simulations by Hopkins \& Quataert (2010a,b).
 The gas is driven inwards successively by a cascade of non-axisymmetries,
first by $m=2$ waves, and then $m=1$. At small scales, clumps and turbulence
can contribute, through dynamical friction, and viscous torques. Gas is indeed
piling up into the center, but intermittently. The time fluctuations
reproduce themselves self-similarly at various scales, in a fractal behaviour. Even when
driven by a bar, the gas flow is intermittent, moderated by the feedback.
 It is then expected that activity periods are also intermittent, 
as observed in the Milky Way for instance.

\begin{figure}[t]
\begin{center}
 \includegraphics[width=13cm]{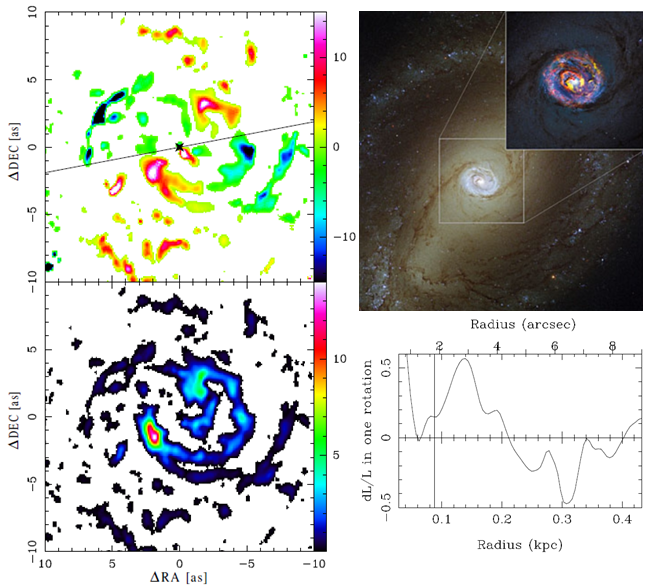}
 \caption{ {\bf Top right:} CO(3-2) ALMA map of the nuclear disk in NGC 1433, superposed
to the optical image (Combes et al. 2013).  The deprojected molecular map is 
reproduced at the {\bf bottom left}, and the derived torque map is at {\bf top left} (Smajic
et al 2014). The torques change sign as expected in a four-quadrant
pattern (or butterfly diagram), following the nuclear bar’s orientation (full line). 
{\bf Bottom right:} relative loss of angular momentum per rotation, versus radius, in the
the galaxy NGC 1433. The gas is stalled in a 200pc ring (the central 50pc 
is perturbed by a molecular ouflow).}
   \label{fig6}
\end{center}
\end{figure}

Feedback can occur also in low-luminosity AGN, like in the Seyfert 2 galaxy NGC 1433
(see Figure \ref{fig6}). The torques inside the nuclear ring are positive, and 
the gas is stalled there, at $\sim$ 100pc scale. Inside, at 10pc scale, a molecular
outflow is detected on the minor axis, corresponding to 7\% of the mass. This is
the smallest outflow rate detected in a nearby galaxy.

\begin{figure}[t]
\begin{center}
 \includegraphics[width=13cm]{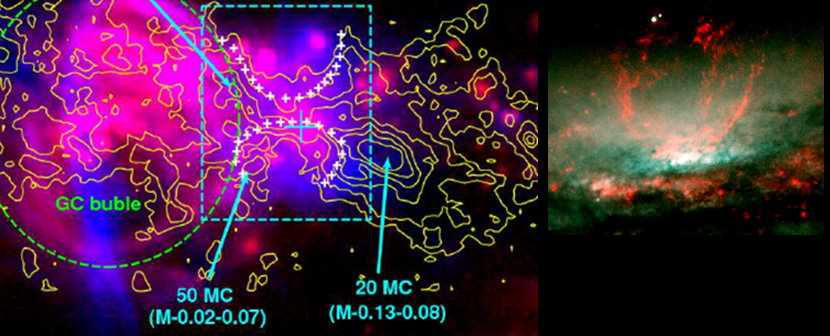}
 \caption{ {\bf Left:} Evidence of a past explosion close to the Galactic center,
with two opposite flows of ionized gas (from Hsieh et al. 2016).
{\bf Right:} Comparison with the tulip-like bipolar outflow in NGC 3079 (Cecil et al. 2001).}
   \label{fig7}
\end{center}
\end{figure}

In the Milky Way, there are several pieces of evidence of past activity, stopped
by feedback. Bubbles have been detected in gamma-rays with Fermi, extending
10 kpc above and below the plane (Su et al 2010). These bubbles correspond also
to synchrotron emission in cm and mm wavelength (WMAP, Finkbeiner 2014). 
Radio emission with Parkes-64m has been detected by Carretti et al. (2013).
These bubbles might have been created by supernovae feedback.
Hsieh et al. (2016) have recently found evidence of a past explosion through
a typical bipolar morphology of the ionized gas (see Figure \ref{fig7}).
The time-scale derived is of 0.5 Myr. This corresponds to the 
hour-glass shape defined by the CS emission, and the polar arc is in
the alignment.

\section{Conclusions}

 The dynamical processes in the Milky Way can be derived by 
analogy with those occuring in nearby galaxies. At large-scale, the
primary bar drives gas inwards from 10kpc to R $\sim$ 100pc.
Then a possible nuclear bar can continue the action from 100pc to 10pc.
 Statistically, gas is driven in about one third of the time, in the life
of the galaxy.

At scales ~1-10pc, other processes,
invoking viscous turbulence, clumps, warps, bending,
  dynamical friction, formation of thick disks/torus, 
will fuel the center, when there is gas. Under the black hole influence,
the $m=1$ instabilities are frequent in nuclear stellar disks.

The gas fueling is moderated by feedback, either from
supernovae or the active nucleus. Activity periods are intermittent,
and a majority of the time, the nuclear region is quenched, as is the Milky way
today. 

The gas ejected by the feedback outside the plane, can fall back
in random orientations. There are frequent
mis-alignment between small scale and large scale disks.
The decoupling of the various scales is also 
expected, due to different dynamical time-scales.

\end{document}